\documentclass[a4paper, 10pt, conference]{ieeeconf}      
\IEEEoverridecommandlockouts                               
\usepackage[style=ieee,backend=biber, url=false, doi=false, isbn=false, maxnames=100]{biblatex}

\usepackage{graphicx} 
\usepackage{epstopdf}

\usepackage{xcolor}
\usepackage{amsmath,array,graphicx}
\usepackage{kantlipsum}
\usepackage{latexsym}
\usepackage{amsfonts}
\usepackage{amssymb}
\usepackage{multirow}
\usepackage{gensymb}
\usepackage{indentfirst}

\title{\LARGE \bf A Case Study of Trust on Autonomous Driving$^{*}$
\thanks{*This research was supported in part by NSF grant CNS-1755784 and Toyota Motor North America ``Digital Twins'' project. }
}

\author{Shili Sheng$^{1}$, Erfan Pakdamanian$^{1}$, Kyungtae Han$^{2}$, \\ BaekGyu Kim$^{2}$, Prashant Tiwari$^{2}$, Inki Kim$^{3}$, Lu Feng$^{1}$
\thanks{$^{1}$Shili Sheng, Erfan Pakdamanian, and Lu Feng are with School of Engineering and Applied Science, University of Virginia, Charlottesville, VA 22904, USA
        {\tt\small \{ss7dr, ep2ca, lf9u\}@virginia.edu}}%
\thanks{$^{2}$Kyungtae Han, BaekGyu Kim, and Prashant Tiwari are with Toyota Motor North America, Mountain View, CA 94043, USA
        {\tt\small \{kyungtae.han, baekgyu.kim, prashant.tiwari\}@toyota.com}}%
\thanks{$^{3}$Inki Kim is with  College of Engineering, University of Illinois at Urbana Champaign, Urbana, IL 61801, USA
        {\tt\small inkikim@illinois.edu}}%
}

\begin{document}
\maketitle
\thispagestyle{empty}
\pagestyle{empty}

\begin{abstract}
As autonomous vehicles have benefited the society, understanding the dynamic change of humans' trust during human-autonomous vehicle interaction can help to improve the safety and performance of autonomous driving.
We designed and conducted a human subjects study involving 19 participants. Each participant was asked to enter their trust level in a Likert scale in real-time during experiments on a driving simulator. We also collected physiological data (e.g., heart rate, pupil size) of participants as complementary indicators of trust. 
We used analysis of variance (ANOVA) and Signal Temporal Logic (STL)  to analyze the experimental data. 
Our results show the influence of different factors (e.g., automation alarms, weather conditions) on trust, and the individual variability in human reaction time and trust change. 
\end{abstract}

\section{Introduction}
Autonomous vehicles  have  achieved  a  high  level of  autonomy  with  the  development  of  various sensors and advanced driver assistance systems.
Although autonomous driving requires less human involvement in the vehicle operation, the trust level can affect humans' interaction with the vehicle and decide humans' reliance on vehicle usage \cite{Korber2018,Bailey2007Automation-inducedTrust}.
Human operators tend to use the automation they trust and reject it when they do not \cite{Pop2015IndividualAutomation}. 
Undertrust can lead to the neglect or under-utilization of automation, while overtrust may cause misuse of automation (e.g., delayed take-over control when human intervention is necessary)~\cite{Parasuraman1997,Lee2004TrustReliance,Lahijanian2016,Miller2016}.
Therefore, it is important to understand the role of trust during human-autonomous vehicle interaction, which can help to improve the safety and performance of autonomous driving. 

Nevertheless, there are many challenges to gain insights into the role of trust during autonomous driving. 
Trust in automation can be influenced by many factors. Intrinsically, a trustworthy autonomous driving system relies on the appropriate integrated implementation of various system components, such as whether to allow manual take-over and how the alarm is delivered. Extrinsically, the ambient environment (e.g., weather conditions) and hazardous incidents (e.g., pedestrian crossing) also introduce uncertainties into the change of trust level. 
Furthermore, as a subjective mind state, humans' trust is difficult to observe and measure. 
The existing studies mostly use post-experiment surveys or questionnaires to evaluate trust level \cite{Desai2012, Rezvani2016, Koo2015}.
However, such methods cannot capture the dynamic change of trust level in real-time~\cite{Setter2016Trust-basedAgents, Xu2015OPTIMo:}, which is important for autonomous driving (e.g., to decide timely driver intervention actions). 
Several recent studies also use physiological data such as electroencephalogram (EEG), galvanic skin response (GSR), gaze tracking, and heart rate variability (HRV) to infer humans' trust and emotional state~\cite{Hu2016,RASTGOO2018}.
But they mostly focus on the trust analysis for a group of participants. 
The group-level  analysis provides a generalized understanding of humans' trust, but  trust can varies from person to person as it is influenced by humans' disposition and past experience. There is a need for the trust analysis of individuals.

In this paper, we present a case study of evaluating humans' trust by a Likert scale in real-time during experiments on a driving simulator. We used analysis of variance (ANOVA) to examine the potential influence of different factors on  trust, including  alarm type, weather conditions, and driving mode. We also examined the corresponding influence of physiological data (e.g., heart rate, pupil size) since they can be complementary indicators of trust.
Furthermore, we used Signal Temporal Logic (STL)  to check patterns of the trust evolution over time, for example, whether  trust decreases when the vehicle automaton gives false alarms, or whether  trust increases when the vehicle performs well. In order to obtain individualized information on how  trust is affected, we also used STL learning technique to optimize the corresponding reaction time constraints for each individual.

The remainder of the paper is organized as follows: Section II summarizes the related work, Section III describes our human subjects study design and ANOVA analysis results, Section IV presents STL analysis results, and Section V draws the conclusions.  

\section{Related Work}
\noindent
\textbf{Factors that affect trust:}  Hoff \textit{et al.}~\cite{Hoff2015} presented a survey of different factors (e.g., system reliability, timing of error, difficulty of error, type of error) that can influence human operators' trust. 
Studies have shown that system reliability can affect the frequency and timing of autonomy mode switch \cite{Desai2012}. Errors in an early stage of automation or on an easy task  have a greater negative impact \cite{Manzey2012HumanAids,Madhavan2006AutomationAids}. 
 Koo \textit{et al.}~\cite{Koo2015WhyPerformance} studied messages that provides different explanations of autonomous driving actions and showed that describing the reason for actions was preferred by drivers and led to better driving performance. In addition, false alarms (i.e., alarms presenting when there is no event) and missing alarms (i.e., no alarm when there is an event) can also affect trust~\cite{Hoff2015,Davenport2010EffectsTask}.    
\noindent
\textbf{Physiological indicators:} 
Hu \textit{et al.}~\cite{Hu2016} built an empirical trust sensor model based on machine learning classification results with EEG and GSR signals. 
Costa \textit{et al.}~\cite{Costa2001TrustEffectiveness} shows that humans' trust is correlated with stress levels, which can be measured by multiple physiological indicators. 
For example, heart rate (HR) and HRV metrics (i.e., the time fluctuation of heart beats) are widely used indicators for stress level \cite{RASTGOO2018,Pereira2017HeartAssessment}. Photoplethysmogram (PPG) signals controlled by the heart's pumping action are used to extract HRV parameters \cite{Elgendi2012OnSignals.}.
In addition, Pedrotti \textit{et al.}~\cite{Pedrotti2014} finds that the pupillary response signal has a good discriminating power for stress detection. Moreover, the pupil diameter increases as the result of sympathetic nervous system activity when a human is under stress \cite{RASTGOO2018}.

\section{Human Subjects Study}
   \begin{figure}[t]
   	\centering
   	\includegraphics[width=3.3in]{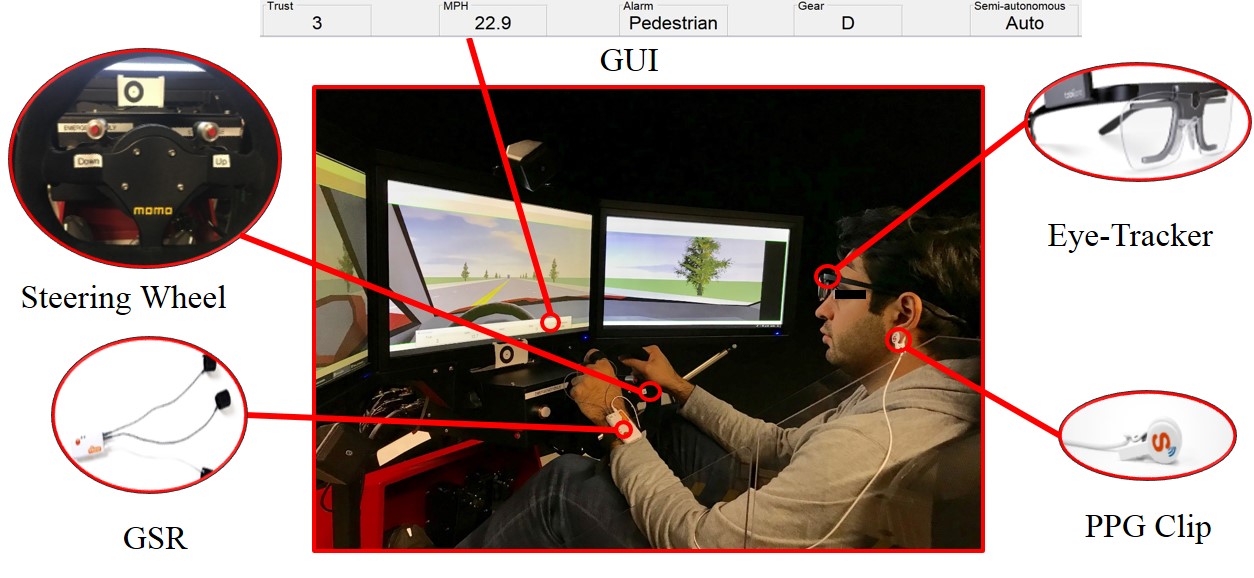}
   	\caption{Participant engaging with simulated driving environment, with GSR, PPG, and eye movement being recorded. The buttons embedded in the steering wheel are used to adjust trust level and switch between manual driving and autonmous driving. The  graphical user interface (GUI) displays the current trust level, the vehicle speed, the alarm detected, the gear, and whether the system is manual driving or autonomous driving.}
   	\label{driver}
   \end{figure} 

\subsection{Driving Testbed Setup}
\begin{figure}[t]
   	\centering
   	\includegraphics[width=3.3in]{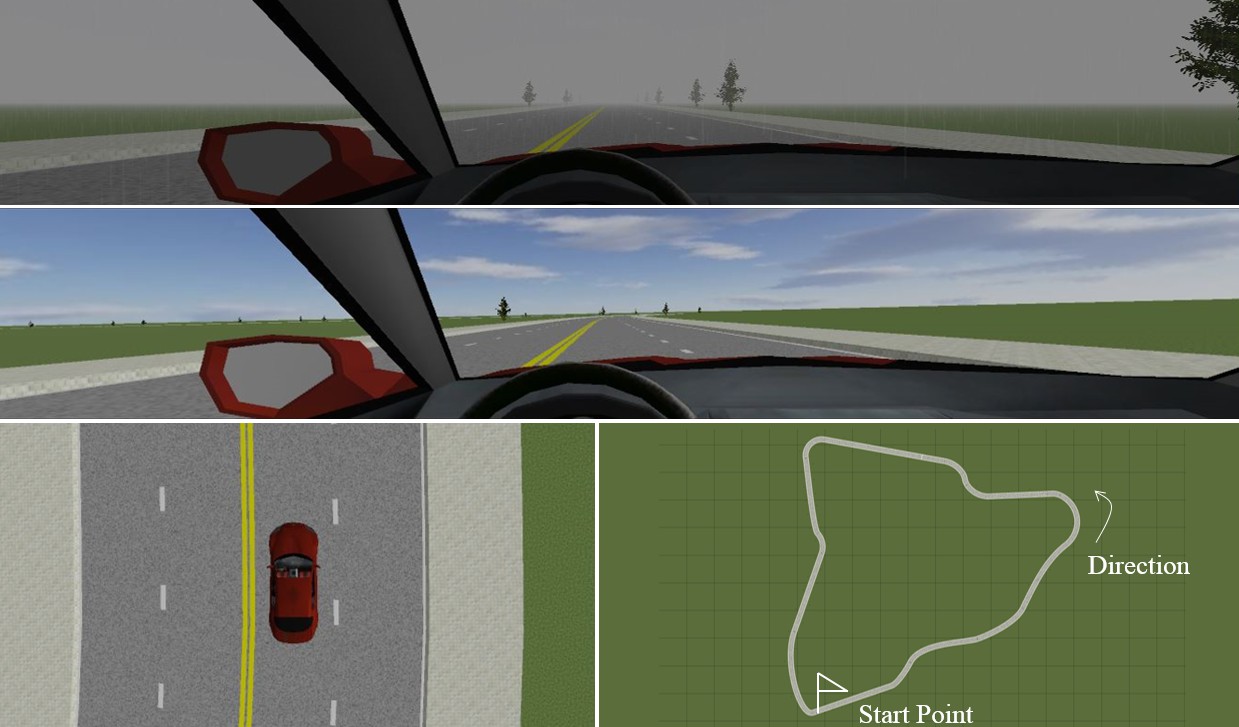}
   	\caption{Scenarios; Top: Driver's view in rainy weather; Center: Driver's view in sunny weather; Bottom-left: Top view of the ego car; Bottom-right: Top view of the scenario, gird spacing: 100 meters.}
   	\label{Scenario}
   \end{figure}

We conducted the experiments in a high-fidelity driving simulator (Force Dynamics 401CR, see Fig.~\ref{driver}), which is a four-axis motion platform that tilts and rotates to simulate the experience of being in a vehicle. The human interacts with the driving simulator through the PreScan software, which can be programmed to simulate autonomous driving scenarios (see Fig.~\ref{Scenario}).
While driving, the participants' physiological data (GSR, PPG, eye-tracking) are collected through the Shimmer3 GSR+ sensor and Tobii Pro Glasses 2. All experimental data are recorded and synchronized via iMotions Biometric Platform~\cite{imotions}.

\subsection{Experimental Design} 

We view trust as \emph{delegation of responsibility for actions to the automation and willingness to accept risk and uncertainty}, following the definition of trust in \cite{Lee2004TrustReliance}.
Our experimental design has one primary dependent variable (i.e., trust), and three independent variables (i.e., alarm type, weather conditions, and driving mode).
We designed 16 driving scenarios considering the different combinations of these variables. 
Each driving scenario contains four hazardous events: (1) a pedestrian crossing the road, (2) an obstacle in front of the lane, (3) a slow-moving cyclist in the same lane, and (4) an oncoming truck from the opposite direction in a nearby lane. At the time of hazard detection, an auditory alarm with a high frequency (750 Hz) went off to alert the driver about the upcoming hazard. We designed four alarm types  (details of each alarm is explained below): \emph{AAAA}, \emph{MMMM}, \emph{FAAAA}, \emph{AAAFA}. All these conditions were counterbalanced so that participants could come across all four alarm types and hazardous events. Trust was evaluated by a 5-point Likert scale (five as the most trust and one as the least trust) in each condition with respect to the following independent variables:
\begin{itemize}
    \item \textbf{Alarm type}: Each driver experienced receiving the following four alarm types randomly in a driving scenario:
    \begin{itemize}
        \item all four alarms were Activated (\emph{AAAA}), 
        \item all four alarms were Missing (\emph{MMMM}), 
        \item an early False alarm (alarm activated for no incident) was triggered  (\emph{FAAAA}),
        \item a False alarm between the third and the fourth incident was triggered (\emph{AAAFA}). \end{itemize}
       
    \item \textbf{Weather conditions}: Two weather conditions (sunny and rainy) were used in this study (see Fig. \ref{Scenario}). In the sunny weather, the driver has clear visibility whereas in the rainy weather, the visibility was set at 240 meters (we assigned the precipitation density as $365,000$ particles per cell). 
    \item \textbf{Driving mode}: 
    Eight of the scenarios were in the fully-autonomous mode and the other eight were in the semi-autonomous mode.  In the fully-autonomous mode, the system stayed in autonomous driving and it would not respond to  participants' operation on the wheel, the brake, or the throttle. On the contrary, in the semi-autonomous mode, the system started in autonomous driving by default, but the participants could switch between autonomous driving and manual driving freely. 
    
\end{itemize}  

 Two buttons embedded in the steering wheel (see Fig. \ref{driver}) were used for two purposes: 1) In the semi-autonomous mode, pressing the two buttons at the same time can switch between autonomous driving and manual driving. 2) When the vehicle is in autonomous driving, pressing the left button once can decrease the trust by one level, and pressing the right button once can increase the trust by one level.  By default, for autonomous driving the trust level is set at three to start with, but the participants can adjust it freely within scale one to five. For manual driving, the trust level is set at zero, and the participants cannot change it.

\subsection{Hypotheses}
Our main hypothesis is that humans' trust is affected by various factors. Missing alarms and false alarms would cause negative impression on humans which lead to a lower trust. People's trust changes due to the variation of weather conditions. In addition, people's trust would be different in the semi-autonomous mode (i.e., drivers can switch between autonomous and manual driving), compared to the fully-autonomous driving mode.
Specifically, we hypothesized that:

\begin{itemize}

    \item \textbf{H1}: 
    Humans' trust is affected by the alarm type. 
    
    \item \textbf{H2}: 
    Humans' trust is affected by the weather conditions.
    
    \item \textbf{H3}: 
    Humans' trust is affected by the driving mode.
    
\end{itemize}
  
\subsection{Experiment Procedure}
Nineteen participants (mean age: 22.57 years, $SD = 3.76$ years, 63\% female) were recruited from the University of Virginia.  The Internal Review Board\footnote{IRB \# 20606: Cognitive Trust in Human-Autonomous Vehicle Interaction} at the University of Virginia has approved the requirements and the study. All of the recruited participants were students ranging in age from 18 to 35 years old. All participants were required to hold a valid driving license with at least one year of driving experience. All participants had normal or corrected-to-normal vision.  

Upon arrival at the lab, the participants were instructed to read and sign an informed consent. The participants were informed that they could quit the experiment at any time without any penalty. They filled out a pre-experiment demographic questionnaire.   
They were instructed to sit in the driving simulator. Overhead lights were turned off. A three-minute baseline experiment was conducted to record GSR, PPG, and pupil diameter. A three-minute training trial was conducted to allow the participants to get familiar with the driving system. Each participant had 16 trials and each trial lasted 180 seconds.
After the experiment, each participant received a \$20 gift card.

\subsection{Data Pre-processing}
GSR are physiological signals captured from the surface of the skin. These signals reflects the electrical conductivity of skin and the arousal of nervous response~\cite{Dawson2011TheDecision-making.}. The average of GSR values, and the average of peaks of GSR values are significantly affected by trust and cognitive load~\cite{Khawaji2015UsingEnvironment}.
We computed GSR peaks from phasic data extracted from GSR signals using a mean filter \cite{dawson2007electrodermal}. For each sample point, the mean GSR of the time interval [-4s; +4s] centered on the current sample was computed. The mean GSR value was subtracted from the current sample. The result is the phasic data. A lowpass filter with cut-off frequency at 5 Hz was applied to phasic data in order to reduce line noise. GSR peaks were found in phasic data between peak onsets ($>0.01\  \mu Siemens$) and offsets ($<0 \ \mu Siemens$).  

Heart rate (HR) and heart rate variability (HRV) are two measures that can vary along with increasing cognitive load \cite{mehler2011comparison}.  The following time domain measures of HRV were calculated from normal-to-normal (NN) of beat-to-beat (R-R interval) variations of consecutive heartbeats \cite{Pereira2017HeartAssessment}. Increasing cognitive load causes decreasing HRV measurements, such as the mean of RR (RRMean), the root mean square successive difference between consecutive NN (RMSSD), the standard deviation of NN intervals (SDNN), and the ratio of adjacent NN intervals differing at least 50 ms (NN50) to the all NN intervals (percentage of NN50 or pNN50).  
In this study, we only relied on pupil size as a  metric  of cognitive load obtain from eye-tracker. Pupil size in millimeters was calculated as the average pupil sizes of both left and right eyes.

\subsection{ANOVA Analysis} 
\renewcommand{\tabcolsep}{2pt}
\begin{table*}[t]
\scriptsize
\centering
\caption{Mean and standard deviations (SD) of dependent variables} 
   \label{SD} 
\begin{tabular}{lll|ll|llll|ll}
\hline
            & \multicolumn{2}{c}{\textbf{Mode}}        & \multicolumn{2}{c}{\textbf{Weather}}    & \multicolumn{4}{c}{\textbf{Alarms}}                                      & \multicolumn{2}{c}{\textbf{Gender}}       \\
            & Fully-Auto     & Semi-auto      & Rainy         & Sunny          & AAAA          & AAAFA         & FAAAA          & MMMM           & Female         & Male            \\ 
\cline{2-11}
\textbf{Heart Rate } &                &                &               &                &               &               &                &                &                &                 \\
 \multicolumn{1}{r}{pNN50}        & 0.26(0.15)     & 0.26(0.16)     & 0.26(0.15)    & 0.26(0.15)     & 0.25(0.15)    & 0.27(0.16)    & 0.24(0.18)     & 0.24(0.15)     & 0.30(0.16)     & 0.22(0.13)      \\
 \multicolumn{1}{r}{RRMean}       & 812.98(109.86) & 803.70(115.88) & 809.12(113.9) & 807.92(112.08) & 802.13(113.7) & 815.18(110.0) & 809.29(112.33) & 806.75(116.95) & 821.74(108.28) & 789.91(116.71)  \\
\multicolumn{1}{r}{RMSSD}      & 52.9(21.02)    & 51.26(21.30)   & 53.38(20.07)  & 51.79(22.31)   & 53.34(24.73)  & 53.40(22)     & 50.8(18.48)    & 50.7(18.99)    & 55.08(19.74)   & 47.95(22.35)    \\
\multicolumn{1}{r}{SDNN}        & 57.0(19.59)    & 57.19(20.7)    & 57.31(20.38)  & 56.88(19.97)   & 57.58(21.8)   & 56.64(18.97)  & 55.76(16.75)   & 58.50(22.71)   & 59.27(19.02)   & 54.11(21.30)    \\
\textbf{Eye-tracker} &                &                &               &                &               &               &                &                &                &                 \\
\multicolumn{1}{r}{Pupil Size}   & 3.87(0.39)     & 4.0(0.39)      & 4.05(0.39)    & 3.82(0.37)     & 3.9(0.41)     & 3.96(0.38)    & 3.9(0.37)      & 3.1(0.42)      & 3.5(0.41)      & 3.8(0.37)       \\
\textbf{GSR  }       &                &                &               &                &               &               &                &                &                &                 \\
\multicolumn{1}{r}{Peaks}       & 13.5(8.8)      & 18.8(11.16)    & 15.2(10.1)    & 16.5(10.7)     & 16.0(10.9)    & 15.78(10.27)  & 16.4(10.57)    & 16.7(6.9)      & 19.7(9.33)     & 11.33(9.8)      \\
\textbf{Trust}       & 3.34(0.81)     & 3.39(0.69)     & 3.13(0.81)    & 3.47(0.7)      & 3.51(0.73)    & 3.44(0.68)    & 3.29(0.72)     & 3.18(0.84)     & 3.24(0.76)     & 3.64(0.79)      \\
\hline
\multicolumn{2}{l}{Note: SD is in parentheses}
\end{tabular}

\end{table*}

Table~\ref{SD} shows statistics (i.e., sample mean and standard deviations) on dependent variables to describe the effectiveness of our  experiments. 
A $4\times 2\times2\times2$ (alarm type [\textit{AAAA}, \textit{MMMM}, \textit{FAAAA}, and \textit{AAAFA}], weather conditions [rainy and sunny],  driving mode [fully- and semi- autonomous],  gender [female and male]) ANOVA with and user ID as a random was undertaken. Tukey HSD tests were used for post-hoc contrasts. Also, a significance level of $0.05$ was used for all
statistical tests, unless stated otherwise.  

Results of a repeated-measures  ANOVA on \textbf{H1} showed significant main effect of alarm type ($F(3,303)=3.11$, $p=0.026$, $\eta^2=0.76$) on average trust. The post-hoc Tukey HSD test revealed that the average trust in \textit{MMMM} was significantly different from \textit{AAAA} and \textit{AAAFA} (see Fig. 3). 
The average trust when all the alarms were missing (\textit{MMMM}) was significantly less than the average trust when all alarms were present  (\textit{AAAA}). Besides, in \textit{MMMM}, the average trust of participants dropped significantly lower than receiving a false alarm in the late stage (\textit{AAAFA}).   

In addition, we ran a repeated-measures ANOVA with the factors of the weather conditions and the driving mode on trust. 
We observed a weak  effect of the weather conditions ($F(1,303)=1.24$, $p=0.26$, $\eta^2=0.03$) and the driving mode ($F(1,303)=0.21$, $p=0.64$, $\eta^2=0.01$) on trust. 
ANOVA also showed significant main effect of gender ($F(1,302)=10.62$, $p=0.001$,  $\eta^2=0.47$)) on average trust. However, we found no statistically significant interaction between alarm type and gender.

Result of ANOVA also revealed that the pupil size, ($F(1,303)=27.29$, $p<0.001$, $\eta^2=0.36$), and the number of peaks in GSR, ($F(1,303)=9.34$, $p<0.003$, $\eta^2=0.41$),  were sensitive to the effect of weather conditions. In addition, results indicated that HRV ($F(1,299)=29.19$, $p<0.001$ , $\eta^2=0.29$) is significantly affected by the gender, as opposed to weather conditions($F(1,303)=0.28$, $p=0.28$, $\eta^2=0.03$) and alarm type ($F(1,303)=0.13$, $p=0.93$, $\eta^2=0.06$). 

\begin{figure}[t]
\begin{center}
  \includegraphics[width=18cm,height=7cm, keepaspectratio]{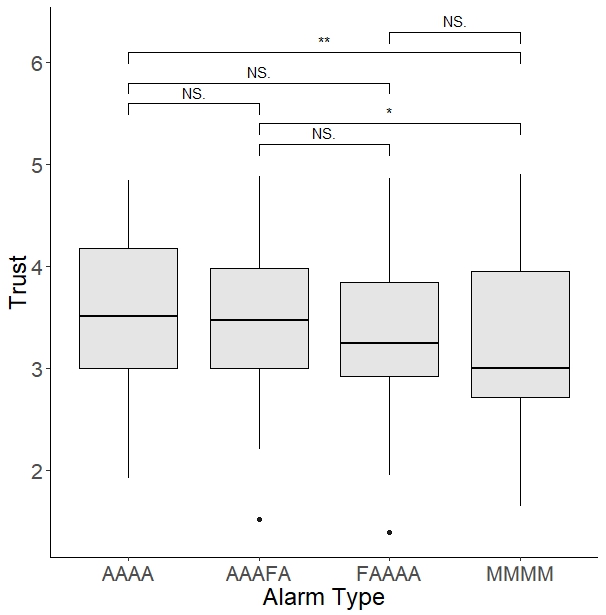}
  \caption{The grouped box plot displays the comparison of the average trust between four alarm types. NS: not significant, \text{**}: $p<0.01$, and \text{*}: $p<0.05$. } 
  \label{fig:overview}
  \end{center}
\end{figure}  

\section{STL-based Analysis}

The result of ANOVA takes the group of participants into consideration, but it lacks the personalized information. Therefore, we employed STL for the further individual analysis.
\subsection{Signal Temporal Logic}
Signal Temporal Logic (STL)~\cite{Maler2004MonitoringSignals} is a formal specification language to express temporal properties over real-values trajectories with dense-time intervals. 
STL is commonly used to describe desired behaviors of cyber-physical systems (e.g., automotive systems, medical devices)~\cite{bartocci2018specification}. 
We used STL to analyze trust signals measured in our human subjects study, in order to identify patterns of trust evolution over time. 

The syntax of a STL formula $\varphi$ over trace $\tau$ is defined as:
$$
\varphi ::= \mu \,\lvert\,  \neg \mu  \,\lvert\,\varphi \wedge \varphi \,\lvert\, \varphi \vee \varphi \,\lvert\, \Box_{[u,v]} \varphi \,\lvert\, \diamondsuit_{[u,v]} \varphi \,\lvert\, \varphi \mathbf{U}_{[u,v]}\varphi
$$
where $[u,v]$ is a closed time interval and  $0\leq u<v<\infty$. The signal predicate $\mu$ is the formula of the form $g(\tau)>0$, where $\tau \in \mathcal{X}$ is a signal variable, and $g$ is a function from $\mathcal{X}$ to $\mathbb{R}$.
The Boolean satisfaction of a given STL formula $\varphi$ is \textit{True} if and only if $\varphi \vDash \mu(\tau(t))$ (i.e., $\mu(\tau(t))>0$). The $\Box$,  $\diamondsuit$, and $\mathbf{U}$ operator stands for "always", "eventually", and "until", respectively. $\tau\vDash \Box_{[u,v]} \varphi$ specifies that $\varphi$ holds at every time step between $u$ and $v$. Similarly, $\tau\vDash \diamondsuit_{[u,v]} \varphi $ specifies that $\varphi$ holds at some time step between $u$ and $v$.  Finally, $\tau\vDash \varphi_{1} \mathbf{U}_{[u,v]}\varphi_{2}$ specifies $\varphi_{1}$ holds at every time step before $\varphi_{2}$ holds, and $\varphi_{2}$ holds at some time step between $u$ and $v$.

\subsection{Checking STL Formulae}
Our data set consists of 19 participants, with 16 trials for each participant. We aggregated each 180-second trial into 1800 rows with a time step of 0.1 second. In order to investigate the dynamic of trust, we calculated the \textit{Trust Change} by subtracting the trust level of the previous second from the one of the current second. 

\begin{table*}[t] \centering 
\caption{STL Formulae} 
   \label{STL Formulae} 
\begin{tabular}{@{\extracolsep{5pt}}llcc} 
\\[-1.8ex]\hline 
\\[-1.8ex]  & Formula  & \#Trial & \#Participant \\ 
\hline \\[-1.8ex] 
\multirow{3}{*}{\text{No Event}}
&$\varphi_{1}\vDash \diamondsuit((\Box_{[0,30]} \text{No Event} )\wedge(\Box_{[0,30]} \text{Autonomous Driving} ) $      &       275  & 19\\
&$\varphi_{2}\vDash \diamondsuit((\Box_{[0,30]} \text{No Event} )\wedge(\Box_{[0,30]} \text{Autonomous Driving} )\wedge\diamondsuit_{[15,16]} \text{Trust Decrease}) $      &       51  & 19\\
&$\varphi_{3}\vDash \diamondsuit((\Box_{[0,30]} \text{No Event} )\wedge(\Box_{[0,30]} \text{Autonomous Driving} )\wedge\diamondsuit_{[15,16]} \text{Trust Increase}) $      &       86 & 19\\
 \hline \\[-1.8ex] 
 
\multirow{6}{*}{Fasle Alarm}
&$\varphi_{4}\vDash \diamondsuit(\text{Early False Alarm} )$ & 76 & 19\\
&$\varphi_{5}\vDash \diamondsuit(\text{Late False Alarm} )$ & 75& 19\\
&$\varphi_{6}\vDash \diamondsuit(\text{Early False Alarm} \wedge\diamondsuit_{[0,10]}\text{Trust Decrease})$ & 17 &9\\
&$\varphi_{7}\vDash \diamondsuit(\text{Late False Alarm} \wedge\diamondsuit_{[0,10]}\text{Trust Decrease})$ & 21&13\\
&$\varphi_{8}\vDash \diamondsuit(\text{Early False Alarm} \wedge\diamondsuit_{[0,10]}\text{Trust Increase}))$ & 11&6\\
&$\varphi_{9}\vDash \diamondsuit(\text{Late False Alarm} \wedge\diamondsuit_{[0,10]}\text{Trust Increase})$ & 17&9\\
 \hline \\[-1.8ex]

 \multirow{6}{*}{Missing Alarm}
&$\varphi_{10}\vDash \diamondsuit(\text{Event Detected} \wedge \text{Alarm Not Activated})$ & 76&19\\
&$\varphi_{11}\vDash \diamondsuit(\text{Event Detected} \wedge \text{Alarm Not Activated} \wedge \diamondsuit_{[0,10]} \text{Trust Decrease})$ & 51&17\\
&$\varphi_{12}\vDash \diamondsuit(\text{Event Detected} \wedge \text{Alarm Not Activated} \wedge \diamondsuit_{[0,10]} \text{Trust Increase})$ & 39&16\\
&$\varphi_{13}\vDash \diamondsuit(\text{Event Detected} \wedge \text{Alarm Activated})$ & 228&19\\
&$\varphi_{14}\vDash \diamondsuit(\text{Event Detected} \wedge \text{Alarm Activated} \wedge \diamondsuit_{[0,10]} \text{Trust Decrease})$ & 117&17\\
&$\varphi_{15}\vDash \diamondsuit(\text{Event Detected} \wedge \text{Alarm Activated} \wedge \diamondsuit_{[0,10]} \text{Trust Increase})$ & 135&19\\

 \hline \\[-1.8ex] 
\end{tabular}
\end{table*}

Breach \cite{Donze2010BreachSystems} is a framework designed for formal analysis and system monitoring. Given a system property as an STL formula, the system is capable of detecting a violation \cite{Watanabe2018RuntimeVehicles}. We used Breach to detect the satisfaction of the formulae. Table \ref{STL Formulae} list all the STL formulae considered: the number of trials that satisfied each formula, and the number of participants who had trials that satisfied the formula.  It should be noted that one trial can satisfy multiple STL formulae.
We analyzed the results of checking STL formulae as follows:

\noindent
\textbf{No Event.}  We used $\varphi_{1}$ to extract the trials in which the participants stayed in autonomous driving and no event is detected for 30 seconds. Then we extracted the trials in which the participants decreased or increased their trust levels at least 15 seconds after the last event (if any) and at least 14 seconds before the next event (if any) by $\varphi_{2}$ and $\varphi_{3}$, respectively. We assume that the change of the participants' trust was not influenced by the events occurring 15 seconds before or 14 seconds after.
In that case, $\varphi_{2}$ stands for the trials in which the participants decreased trust during only lane keeping driving. The results of $\varphi_{2}$ and $\varphi_{3}$ demonstrate that trust is more likely to increase after a period of driving without dealing with any events.

\noindent
\textbf{False Alarm.} The results of $\varphi_{4}$ and $\varphi_{5}$ show the number of trials in which participants encountered an early false alarm and a late false alarm, respectively. As described in the experimental design in Section II, there were 76 trials with early false alarm and 76 trials with late false alarm. However, in one trial, the participant accidentally drove off the road and avoided the area where the late false alarm was designed. The results of $\varphi_{6}$ and $\varphi_{7}$ show the trials where  trust decreased within 10 seconds after a false alarm occurring. The results of $\varphi_{8}$ and $\varphi_{9}$ show the trials where trust increased within 10 seconds after a false alarm occurring. Of the early false alarms, 22.4\% caused trust to decrease, while 28.0\% of late false alarms caused trust to decrease. Of the early false alarms, 14.5\% caused trust to increase, while 22.7\% late false alarms caused trust to increase. In total, 25.2\% of false alarms caused to trust to decrease, while 18.5\% caused trust to increase. 

\noindent
\textbf{Missing Alarm.}    
We extracted the trials when the participants decreased and increased trust level within 10 seconds of a missing alarm event by using $\varphi_{11}$ and $\varphi_{12}$, respectively. The results show that the probability of trust decreasing after a missing alarm event is 67.1\% while the probability of it after an activated alarm event is 51.3\%. On the contrary, the results show that the probability of trust increasing after a missing alarm event is 51.3\% while the probability of it after an activated alarm event is 59.2\%. In other words, the scenarios with alarms not being activated have a greater negative impact on the trust level compared to the scenarios with activated alarms.  

\subsection{Learning Individualized Parameters}
The upper bound time we used in the STL formuale $\varphi_{6}$, $\varphi_{7}$, $\varphi_{8}$, and $\varphi_{9}$ are 10 seconds. We assume that participants can react to change trust levels within 10 seconds in general. In fact, some participants reacted faster than others. To obtain a tighter reaction timing bound for each participant, we used the Temporal Logic Extractor (TeLEx) tool \cite{jha2017telex} to learn the optimal STL parameter from each participant's data. TeLEX takes the input of parametric STL formulae and each participant's trial data, and outputs the learned parameter value of reaction timing bound for each participant. 

\begin{table}[t] \centering 
\centering
\caption{Reaction time (seconds)} 
   \label{TableLearning} 
\begin{tabular}{ccccccc} 
\hline
\multicolumn{1}{l}{Participant} & \multicolumn{1}{l}{$\varphi_{6}$} & \multicolumn{1}{l}{$\varphi_{7}$} & \multicolumn{1}{l}{$\varphi_{8}$} & \multicolumn{1}{l}{$\varphi_{9}$} & \multicolumn{1}{l}{Mean}& \multicolumn{1}{l}{SD} \\ 
\hline
A                               & 4.2                        & 3.5                       & 1.0                          & 1.7                       & 2.6         & 1.49               \\
B                              & 4.5                        & 5.5                       & 3.9                        & 5.1                       & 4.75             &0.7            \\
C                             & 2.7                        & 6.5                       & 7.2                        & 9.5                       & 6.475             &2.82            \\
D                              & 3.5                        & 3.7                       & 8.9                        & 9.1                       & 6.3              &3.12           \\
\hline
\end{tabular}
\end{table}

Table~\ref{TableLearning} shows the parameters for four participants. 
Participant A used shorter time to increase trust with respect to the early false alarm ($\varphi_{8}$) and the late false alarm ($\varphi_{9}$) than participants C and D. Participant B had the smallest standard deviation of reaction time, potentially due to more focused engagement in the driving. The results in Table~\ref{TableLearning} demonstrate that there is individual variability in human reaction time and trust change. However, the learned human reaction time may not be accurate due to the small data sample (i.e., only 25 unique trials were found to satisfy these four formulae for these four participants). Further experiments would be needed in order to estimate more accurate individual reaction time.

\section{Discussion and Conclusion}
 This paper used ANOVA and STL to analyze the results in universal and detailed perspective, respectively. ANOVA does not provide the details such as whether false alarm will cause the trust to change. In addition, individual's trust varies from person to person. The result of ANOVA takes the group of participants into consideration but it lacks the personalized information.   

STL framework, on the other hand, allows us to study the user-specific detail of the driving session and output individualized pattern. STL formulae have a mathematically succinct form and can be defined to detect satisfaction or violation behavior of the driving system. With STL parameter synthesis to determine the optimal parameters, STL formulae can be generated to fit individual pattern. 
The structure of the formulae relies on the domain knowledge of the system designer. More research is required to learn new knowledge of the system directly from the observed data. Furthermore, the analysis of physiological data typically requires pre-processing. The direct STL monitoring on physiological signals could be an extension exploration. We demonstrated that STL learning approach can be used to infer individual reaction time. However, we would need to conduct further experiments in order to learn more accurate parameter values. 

The participants in this study are mostly college students with engineering backgrounds. A population with more diversity in age and knowledge background can contribute to more generalized analysis.

In conclusion, this paper presents a case study for trust on autonomous driving. We used ANOVA and STL  to examine how possible factors affect humans' trust change. 
Our ANOVA results show that the missing alarms have a significant impact on humans' trust while driving mode and weather conditions do not. 
We also did ANOVA analysis on physiological data as complementary indicators of trust. It shows that pupil size and number of peaks in GSR were sensitive to the effect of weather. In future work, we plan to explore the relationship between humans' trust and the physiological data.
The results of STL analysis show the variability in human reaction time and trust change.

\printbibliography
\end{document}